\documentclass[preprint,12pt]{elsarticle}

\usepackage{amssymb}

\journal{Physics of the Dark Universe}

\begin{document}

\begin{frontmatter}



\title{Overview of Non-Liquid Noble Direct Detection Dark Matter Experiments}


\author{J. Cooley}

\address{Department of Physics, Southern Methodist University, Dallas, TX  75275, USA}

\begin{abstract}
In the last few years many advances have been made in the field of dark matter direct detection.  In this article I will review the progress and status of experiments that employ detection techniques that do not use noble liquids.  First, I will give an introduction to the field of dark matter and discuss the background challenges that confront all dark matter experiments.  I will also discuss various detection techniques employed by the current generation and the next generation of dark matter experiments.  Finally, I will discuss recent results and the status of current and future direct detection experiments. 
\end{abstract}

\begin{keyword}



\end{keyword}

\end{frontmatter}


\section{Introduction}
The revolution in precision cosmology of the last decade has revealed conclusively that about a quarter of our universe consists of dark matter~\cite{Ade:2013zuv, Bertone:2004pz}. Despite the abundant evidence for the existence of dark matter, its constituents have eluded detection.  

Phenomenology at the intersection of particle physics, cosmology and astrophysics gives well-motivated candidates for dark matter. Dark matter candidates naturally arise from theories that explain the radiative stability of the weak scale. These theories include supersymmetry~\cite{Jungman:1995df} and the existence of extra dimensions~\cite{Agashe:2004ci, Cheng:2002ej, Servant:2002hb}. Candidates from these theories are representative of a generic class of weakly interacting massive particles (WIMPs). In the hot early universe WIMPs would have been in thermal equilibrium. They would have decoupled as the universe expanded and cooled, leaving behind a Òrelic abundanceÓ of these particles in the universe today. In order to explain the observed relic abundance, the WIMP interaction cross section must be consistent with the weak scale~\cite{Kolb:206230}.

If dark matter is composed of WIMPs, their existence in the local galactic halo could be detected through their elastic scattering off nuclei in a terrestrial detector~\cite{PhysRevD.31.3059}.  The expected interaction rate is quite low, less than 0.01 event/(kg$\cdot$day)~\cite{Engel:1992bf}, much lower than the radioactive background of most materials. Hence, any detector would need to be housed deep underground to protect it from cosmic rays and fabricated using radioactivity-free detector materials.

\section{Backgrounds}
Dark matter experiments are faced with a variety of background sources including products of cosmic ray interactions and radioactivity from the environment and detector materials. Although the specifics of dealing with these backgrounds vary from one experiment to another, the general techniques used to reduce these sources are the same.

Dark matter detectors are sited in deep underground laboratories in order to reduce the number of neutrons produced by cosmic ray muon interactions.  As an illustration of the effect that substantial depth can have consider that the neutron rate from cosmic rays at 15 meters water equivalent (mwe) is approximately 1/(kg$\cdot$s). However, at 2000 mwe that rate reduces to approximately 1/(kg$\cdot$day).  Another technique employed by dark matter experiments to reject these neutrons is to surround the detectors with active muon vetoes made of either scintillating panels or water shields.  These vetoes can be used to reject neutron events resulting from muon interactions.

Neutrons from natural radioactivity are another background that dark matter experiments frequently encounter~\cite{Mei2009651}. These neutrons result from ($\alpha$, n) reactions and spontaneous fission of uranium and thorium nuclides that are natural in all geological formations. An ($\alpha$, n) reaction occurs when uranium or thorium decays giving off an $\alpha$-particle which can react with elements such as Al or Na producing a neutron. The number and energy of neutrons produced from these reactions varies depending on the type of material. Passive shielding made from hydrocarbons, such as polyethylene, is typically used to reduce these backgrounds.

Another source of backgrounds is intrinsic to the detector target.  These intrinsic backgrounds include decays of $^{40}$K, $^{210}$Pb and $^{14}$C that can mimic the WIMP signature in dark matter detectors.  Of these contaminants, $^{210}$Pb is of particular concern as it has a half life of 22.3 years and upon decay, emits a beta particle that can mimic a dark matter signature.  $^{210}$Pb is a decay product of $^{222}$Rn which is airborne and has a half-life of 3.8 days.  To reduce backgrounds from $^{210}$Pb, precautions are taken during detector storage, fabrication and deployment to minimize exposure of materials to radon.  In addition, experiments have a variety of analysis techniques that are used to discriminate these beta events from a dark matter signal event.  These techniques yield background leakage rates of $<$ 1.7 $\times$ 10$^{-5}$ at the 90\% confidence level~\cite{Agnese:2013ixa}.

Finally, photons, which can come from environmental radioactivity or radiative contaminants in the detector itself or its shielding, may also contribute to the background seen by dark matter detectors. To reduce this type of background experiments frequently use passive shielding composed of materials such as ancient lead and copper. In an effort to further reduce the photon and neutron backgrounds, the detectors and shielding are built using low-radioactivity materials~\cite{Loach:2013ap}.

Although not a problem for the current generation of dark matter detectors, a future background of concern is neutrinos from astrophysical sources.  Events from coherent neutrino scattering can mimic the WIMP signal~\cite{Billard:2013qya}.  Sources of neutrino conerent scattering include $^{8}$B solar neutrinos at low masses (~10 GeV) and atmospheric and diffuse supernova neutrinos at high masses (~100 GeV).

\section{Detection Techniques}
The current experiments employ one of two strategies for detecting dark matter. 
The first strategy is to look for a variation in the total event rate~\cite{PhysRevD.33.3495, PhysRevD.37.3388} or event direction~\cite{PhysRevD.37.1353, Copi199943} that is expected due to Earth's motion through the galaxy.
The difficulty with this method is that one looks for a small, time-varying signal on top of a huge background. It requires a large detector volume and a stable, nonfluctuating background. The advantage is one does not need to know the characteristics of the background as long as it is not time-varying. 

The other strategy is to substantially reduce the background to near zero and detect WIMP-nucleon interactions. The difficulty with this approach is characterizing the backgrounds and achieving an extremely low-background experimental environment. 

Globally there are a variety of detector mediums used including crystals such as Ge, Si, CaWO$_{4}$, NaI and CsI, superheated fluids and gels such as CF$_{3}$ and C$_{4}$F$_{10}$ and liquid noble gasses such as Xe and Ar. 
There are three observables that result from the interactions of WIMPs in these mediums:
ionization, phonons resulting from the interaction of the WIMP with a nucleon in the crystal lattice, and scintillation.  Experiments composed of these mediums use one or two of these observables to differentiate candidate signal events from the ubiquitous background.  

Dark matter candidates and neutrons are assumed to interact with the targetÕs nucleus producing nuclear recoils while most background particles ($\gamma$-rays and electrons) interact with electrons surrounding the nucleus producing electron recoils. Experiments that use two observables have excellent discrimination between candidate signal events and backgrounds can be obtained from the fractional energy (i.e. ionization energy divided by phonon energy).   This quantity is known as 'yield.'

Another class of experiments uses superheated bubbles or gels to observe the interactions of WIMPs in the target medium.  Mediums used by these experiments include CF$_{3}$I and C$_{4}$F$_{10}$.  This technique is discussed in further detail in Section ~\ref{superfluids}.

It is important to note that no single technology can tell us everything we want to know about dark matter.   It is necessary to have several technologies in different locations and it is important to have complimentary technologies to probe both the high-mass and low-mass regions and to probe both spin-independent and spin-dependent WIMP-nucleon scattering to maximize discovery potential.  If any experiment does observe dark matter, the next objective will be to identify its particle nature.  In order to do so, it will be important to have information about both its spin-independent and spin-dependent couplings and to determine the mass of the dark matter particle, it will be necessary to measure the interaction rate in multiple targets.

\section{The Low-Mass Region}
Light or low-mass dark matter candidates are those whose masses are smaller than 10 GeV/c$^{2}$.  The motivation for looking for candidates of this type stems from three reasons.  First, no signal has thus far been detected by direct detection experiments looking at higher masses or at the LHC.  Second, many particle physics models exist that provide candidates for light dark matter.  These models include supersymmetry, asymmetric dark matter and others~\cite{PhysRevD.80.035008, Essig:2010ye, PhysRevD.79.115016, Cerdeno:2014cda}.  Finally, several experiments have published the existence of small excesses over their predicted background in the low-mass region~\cite{Agnese:2013rvf, Aalseth:2012if, Angloher:2011uu, Bernabei:2010mq}.  Although there is no concrete evidence that these experiments are observing dark matter, it is imperative that current and future experiments explore this largely unexplored region.

\subsection{CoGeNT}
The CoGeNT~\cite{Aalseth:2012if} experiment uses a p-type point-contact germanium detector of mass 440 g to detect the passage of dark matter particles.  The detector acts as an ionization spectrometer. It collected data in the Soudan Underground Laboratory in Minnesota, USA  from Dec. 4, 2010 to March 6, 2011 when data collection stopped due to a fire in the mine.  Data collection resumed again in July 2011.  

This experiment reported an exponential excess of events at low energies after subtracting of known backgrounds due to cosmogenic activation of the germanium detector.  They used 442 live-days spanning the period from December 4, 2010 to March 6, 2011.  Recently, the collaboration performed a maximum likelihood analysis on 3.4 years of CoGeNT data.  This new analysis includes data taken through April 23, 2013 and observes an excess of events over predicted background.  Taking into account systematic uncertainties in the probability density functions (PDFs) used to model both the dark matter signal rates and the background rates, the significance of the signal hypothesis is less than 3$\sigma$ ~\cite{Aalseth:2014jpa}.

CoGeNT continues to take data.  The design of a larger experiment called C-4 is underway.  This experiment aims to have a total mass ten times larger than CoGeNT, a decrease of a factor of 20 in backgrounds, and  lower threshold.  

\subsection{MALBEK}

MALBEK (MAJORANA Low-background BEGe detector at KURF) is a 450 g Canberra Broad Energy Ge (BEGe) detector located at a depth of 1450 mwe in the Kimballton Underground Research Facility (KURF) in Virginia.  This detector was built as part of the MAJORANA demonstrator program whose goal is to demonstrate backgrounds low enough to justify building a ton-scale 0$\nu \beta \beta$ decay experiment.  However, this detector, which is similar to the CoGeNT detector, can also be used to perform a search for light WIMP dark matter and directly test the CoGeNT signal region~\cite{1742-6596-375-1-012014}.  Preliminary results from a 221 day data run are in tension with the dark matter interpretation of the CoGeNT excess at the 90$\%$ confidence level~\cite{Giovanetti:2014lxa}.

\subsection{CRESST}
The CRESST II experiment is located in the Laboratori Nazionali de Gran Sasso, Italy at a depth of approximately 3500 mwe.  The detector consists of eight CaWO$_{4}$ crystals instrumented to readout phonon energy and scintillation.  Each crystal has a mass of ~300 g and is operated at a temperature of ~10 mK.  Discrimination between electron recoil and nuclear recoil events comes from light yield, the ratio of scintillation energy to phonon energy.  The dominant background from radioactivity produces electron recoils.

Results were obtained from a net exposure of 730 kg-day (June 2009 to March 2011).  In that time 67 events were observed in the acceptance region.  The analysis used a maximum likelihood in which two regions favored a WIMP signal in addition to known backgrounds~\cite{Angloher:2011uu}.  As these results had a large number of events, the CRESST II experiment was upgraded  to increase the detector mass and reduce the considerable number of background events.  A new clamp design is anticipated to reduce the number of alpha events and additional shielding is expected to reduce the overall number of events.  Recent results from a single upgraded detector probed new spin-independent WIMP-nucleon parameter space.  The excess of events observed over predicted background in the previous analysis was not confirmed~\cite{CRESST:2014}.

\subsection{CDMS}
The CDMS II experiment operated in the Soudan Underground Laboratory from June 2006 until March 2009.  The experiment consisted of 19 Ge (~239 g each) and 11 Si (~106 g each) detectors that were instrumented to readout the phonon and ionization energy of an event.  Discrimination between electron and nuclear recoils comes from ionization yield, the ratio of ionization to phonon energy.  Electron recoils occurring within the first ~10 $\mu$m of the detector surface can exhibit reduced ionization collection and appear as nuclear recoils.  However, these events can be identified via phonon pulse shape discrimination providing a combined misidentification rate for electron recoils of less than 1 in 10$^{6}$.

The most recent analysis of CDMS II data included 8 Si detectors operated between July 2007 and September 2009.  Three events were observed in an exposure of 140.23 kg-days over a predicated background of ~0.5 events.  A profile likelihood analysis favored a WIMP + background hypothesis over the known background estimate as the source of the excess at the 99.81$\%$ C.L.  The maximum likelihood occurs at a WIMP mass of 8.6 GeV/c$^{2}$ and a spin-independent WIMP-nucleon cross section of 1.9 $\times$ 10$^{-41}$ cm$^{2}$~\cite{Agnese:2013rvf}. 

\subsection{SuperCDMS} 

The SuperCDMS collaboration is currently operating five towers of advanced iZIP detectors in the existing cryostat located in the Soudan Underground Laboratory in Minnesota, USA.  Similar to the CDMS~II detectors, each iZIP is instrumented to readout phonon and ionization energy and has the ability to discriminate between electron recoils and nuclear recoils using ionization yield and pulse shape.  The iZIP detectors differ from their predecessors in that the ionization and phonon sensors are interleaved on the top face and bottom face of each detector.  This allows the surface event background to be identified by the asymmetry of their charge collection.  Events occurring near the surface will generate a charge signal on only one face of the detector while events occurring in the bulk will generate a charge signal on both detector faces.


In order to test the CDMS Si results and to further explore WIMP masses lower than 10 GeV/c$^{2}$, data were collected using one iZIP detector operated in a new mode, CDMSlite.  By increasing the detector voltage to 69 V, CDMSlite leverages Neganov-Luke amplification to obtain low detector threshold with high resolution~\cite{Luke1990406, neganov, wang}.  The disadvantage of this mode is that the experiment measures ionization only and gives up the excellent event-by-event discrimination of nuclear recoils.  Using this method, the SuperCDMS collaboration was able to set an upper limit that excluded some of the favored region of the CoGeNT and CDMS Si analyses under the assumptions of spin-independent WIMP-nucleon interactions using the standard halo model~\cite{Agnese:2013jaa}.

Recently the SuperCDMS collaboration released new results from a low-threshold analysis based on 557 kg-days of data obtained between March 2012 - July 2013.  This analysis used a background model that was developed using simulations based on sideband and calibration data.  It employed a boosted decision tree to discriminate between events resulting from a potential signal and those resulting from backgrounds.   There were 11 candidate events observed on an expected background of 6.2$^{+1.1}_{-0.8}$ events~\cite{Agnese:2014aze}.  This resulted in an upper limit on the spin-independent WIMP-nucleon interactions that disfavors  the interpretation of events seen by CoGeNT and CDMS (Si) as a WIMP signal assuming standard WIMP interactions and a standard halo model.

The SuperCDMS experiment at Soudan continues to take data.  Plans are well underway for a 110 kg SuperCDMS SNOLAB experiment.  The baseline design calls for a combination of Ge and Si detectors that would be operated in both the standard SuperCDMS mode and the new CDMSlite mode.

\subsection{EDELWEISS}

The EDELWEISS-II experiment was located in the Laboratoire Souterrain de Moudan (LSM) located between Italy and France at a depth of 4800 mwe.  Similar to SuperCDMS, EDELWEISS II uses interleaved detectors that measure both phonons and ionization.  However, instead of using TESs to measure phonon energy, EDELWEISS-II measures heat using a neutron transmutation doped (NTD) Ge thermal sensor.  

Recently, the EDELWEISS collaboration conducted a search for WIMPs with masses less than 20 keV using data obtained from four of their interdigitated detectors during a 14 month period in 2009 and 2010.  They observed one event in 113 kg-days of exposure.  This yielded an upper limit of 1.0 $\times$ 10$^{-41}$ cm$^{2}$ for a WIMP of mass 10 GeV~\cite{Armengaud:2012pfa}.

Commissioning and construction of  EDELWEISS-III, a 24 kg fiducial mass experiment, is underway.  Several improvements have been made over EDELWEISS-II including a lower detector threshold and improved resolution.  Special attention was made to reduce backgrounds through additional shielding and improved radiopurity in several detector components.  Finally, improvements are also being made in the discrimination of background events, including surface events.  

\section{Annual Modulation Analyses}

An annual modulation in the differential WIMP event rate occurs due to the Earth's orbit around the sun~\cite{PhysRevD.33.3495, PhysRevD.37.3388}.  Earth's speed with respect to the galactic rest frame is a superposition of the motion of Earth around the Sun and the Sun's rotation around the galactic center.  This speed is largest in the summer when the component of the Earth's orbital velocity vector in the direction of solar motion is the largest.  As a result, the differential WIMP rate has an annual modulation.  The time of year at which the modulation peak will occur depends on the recoil energy of the WIMP~\cite{Primack:1988}.  This effect can be exploited to search for dark matter.

\subsection{DAMA/LIBRA}
The DAMA/LIBRA experiment is located in the Laboratori Nazionali de Gran Sasso, Italy
at a depth of approximately 3500 mwe.  The DAMA/NaI experiment was an 100 kg array of
scintillating NaI crystals operated from 1996 to 2002.  The DAMA/LIBRA experiment is an upgrade to an array of
250 kg scintillating NaI crystals which began operation in 2003.  
This collaboration has observed a modulation of events seen in their experiment consistent with that expected from the WIMP halo for over 13 cycles~\cite{Bernabei:2010mq}.  The signal is only seen in the experiment's lowest energy bin.  However, its significance is 8.9 $\sigma$.  An interpretation of these results as WIMP-nucleon interactions is in strong contention with results from other dark matter experiments.

\subsection{KIMS}
The Korea Invisible Mass Search (KIMS) experiment is located in the South Korean Yangyang Underground Laboratory at a depth of 700 m .  The experiment consists of an array of 12 CsI(T$\ell$) scintillating crystals with a total mass of 103.4 kg.   The crystals are installed in a copper shield.  Unlike DAMA/LIBRA, the KIMS experiment uses pulse shape discrimination to distinguish between events from nuclear recoils and events from electron recoils.  Their most recent analysis had an exposure of 24,524.3 kg-days and resulted in the observed number of nuclear recoil events consistent with zero~\cite{PhysRevLett.108.181301}.  Analysis of the 3.6-5.8 keV energy range, which corresponds to the 2 - 4 keV energy range in NaI(T$\ell$), results in an upper limit at the 90$\%$C.L.  of 0.0083 counts/day/kg/keV, well below the DAMA/LIBRA signal amplitude.

The KIMS collaboration has many plans to upgrade their experiment.  The first of these efforts is KIMS-CsI which is an upgrade of the current detectors.  The plan is to lower the detector threshold to ~1.5 keV.  The second effort is the KIMS-NaI experiment.  The collaboration is trying to duplicate the DAMA/LIBRA experiment using 200 kg of NaI(T$\ell$) crystals.  This experiment is slated to run in 2015-2016.  Finally, plans are underway for the KIMS-LT experiment.  This experiment would use scintillating bolometers such as $^{40}$Ca$^{100}$MoO$_{4}$ crystals to improve sensitivity to low mass dark matter particles.  That experiment is currently slated for operation in 2019-2022.

\subsection{CDMS II and CoGeNT}
The CoGeNT collaboration recently released results of an annual modulation analysis of their data taken over a three year period in the Soudan Underground Laboratory, Minnesota.  Their analysis yields a preference for a modulation over the null hypothesis at a level of 2.2$\sigma$.  Although the phase of the modulation seemed to be consistent with DAMA/LIBRA, the observed modulation amplitude was found to be 4 - 7 times larger than that predicted by the Standard Halo Model~\cite{CoGeNT:2014}

The CDMS II collaboration has also performed an annual modulation analysis on data taken from October 2006 to September 2008.  This experiment is also sited in the Soudan Underground Laboratory.  This analysis ruled out modulated rates with amplitudes greater than 0.07 [keV$_{nr}$ kg day]$^{-1}$ in the 5.0 - 11.0 keV$_{nr}$ energy region~\cite{CDMSII:2012}.  This is inconsistent with the modulation seen by CoGeNT.

\subsection{Future Programs:  ANAIS, DM-Ice and SABRE}
The ANAIS collaboration plans to build a 250 kg array of NaI(T$\ell$) detectors to study the annual modulation effect seen by the DAMA/LIBRA experiment~\cite{2014arXiv1404.3564A}.  This experiment will be housed in the Canfranc Underground Laboratory (LSC) in Spain.  Currently, the collaboration has 2 prototype detectors (ANAIS-25) taking data at LSC.  Preliminary results indicate excellent light efficiency of 12-16 phe/keV.  However, the  $^{40}$K bulk NaI content is 41.7 $\pm$ 3.7  ppb, much higher than DAMA/LIBRA.  They are working to reduce their $^{210}$Pb activity and they continue to struggle with contamination of their signal region by cosmic activation of their crystals.  Work is underway to develop low energy event selection.

The DM-Ice17 experiment was deployed at a depth of 2200 meters water equivalent at the South Pole in December 2010.  This detector is designed to test the feasibility of operating a low-background NaI experiment at the South Pole to directly test the annual modulation WIMP signal observed by the DAMA/LIBRA experiment.  Since DM-Ice is sited in the Southern Hemisphere, it will be able to distinguish any modulation resulting from seasonal effects that have reversed phases between the Northern and Southern Hemispheres from a dark matter modulation which is constant in phase between the Hemispheres.  The detector consists of two 8.47 kg NaI(Tl) scintillating crystals obtained from the NAIAD experiment~\cite{Ahmed:2003su}.  Each crystal along with two photomultiplier tubes, light guides, data acquisition and control electronics, is housed in a stainless steel pressure vessel.  Data recently released from the DM-Ice17 collaboration demonstrated the feasibility of remote operation and calibration and stable environmental conditions with a detector livetime of 99$\%$.  The background observed by the experiment was consistent with simulations of expected contaminates~\cite{Cherwinka:2014xta}. 

The SABRE (Sodium-iodide with Active Background Rejection) experiment will also use a NaI target to study the annual modulation observed by DAMA/LIBRA.  One background of particular concern has been from the decay of $^{40}$K.  This decay most frequently happens through $\beta^{-}$ decay to the ground state $^{40}$Ca or to $^{40}$Ar though electron capture to the excited state $^{40}$Ar* and the subsequent emission of a gamma ray.  Positron emission to the ground state of $^{40}$Ar is also possible although its branching fraction is very small.  Finally, it is possible, although not measured, for $^{40}$K to decay directly to the $^{40}$Ar ground state through electron capture.  The energy in this final decay is primarily carried away by a neutrino, with 3 keV of energy observable from Auger electron emission and X-ray yield upon K-shell capture~\cite{Pradler:2012qt}.  This background is poorly understood and is important to any dark matter interpretation of the DAMA/LIBRA result. 

The SABRE experiment plans to address this issue through the use an active veto to drastically reduce the number of gamma background events.  The veto will not only reject gamma events due to $^{40}$K contamination in the crystals, but also external backgrounds and those from the experimental components.  The experiment is currently in the building phase.

\section{Directional Searches and Techniques}

Another technique exploits the direction of the recoil from a WIMP-nucleon interaction in the detector medium.  A detector located at 45 degree latitude on Earth will see the dark matter wind oscillate in direction over the course of a day~\cite{PhysRevD.37.1353, Copi199943}.  This is a sidereal, not diurnal, effect.  The use of a low pressure gas as a target microscopically extends the particle tracks.   Background rejection then comes from measuring the direction, or head-to-tail effect, of a particle as it passes through the target.  If the particle comes from the direction of the WIMP wind, it is deemed a dark matter candidate, if it comes from any other direction, it is identified as background.

\subsection{DRIFT}
The DRIFT collaboration is currently operating an 800 L fiducial volume (DRIFT-II) in the Boubly facility in the UK~\cite{Daw:2013waa}.  The detector is a TPC with MWPC wire readout.  As a target it uses two low pressure gasses:  CS$_{2}$ operated at 30 Torr and CF$_{4}$ operated at 10 Torr and can probe the WIMP - $^{19}$F spin-dependent cross section.  Discrimination of dark matter signals from background comes both from directional information and track length.  Alpha particles, a background of concern, tend to have track lengths that are $\sim$100s of mm long while recoils tend to have track lengths on the order of $\sim$1 mm.  The background of primary concern at the present time is alpha particles coming from the cathode.  An on-going background reduction program aims to reduce these backgrounds through the reduction of $^{222}$Rn and better fiducialization.  

A second DRIFT-II module is currently being constructed and engineering for a proposed 8 m$^{3}$ experiment called DRIFT-III is underway.

\subsection{DMTPC}
The DMTPC detector is currently in the prototyping phase with a goal to design a m$^{3}$-scale detector that could be replicated for large target mass~\cite{Lopez:2013ah}.  The detector is a TPC with CCD, charge and PMT readout.  It uses low-pressure CF${_4}$ gas at ~50 Torr and can probe the WIMP - $^{19}$F spin-dependent cross section.  In addition to discrimination via the head-to-tail effect, electron recoils can often be identified by their low ionization density.  Currently there are prototypes running at the Waste Isolation Pilot Plant (WIPP) in Carlsbad, New Mexico and at MIT.

\section{Superheated fluids}\label{superfluids}
Superheated fluids provide an alternative to the traditional detectors that rely on discrimination between electron and nuclear recoils to distinguish background events from dark matter signals.  By appropriately choosing the operating temperature and pressure, the detectors using superheated fluids become blind to electron recoils.  When a particle interaction in a chamber filled with a superheated fluid in a metastable state deposits energy above a threshold in a small enough radius, an expanding bubble is formed.  Smaller or more diffuse energy depositions will result in a bubble that immediately collapses.  

\subsection{COUPP}
The COUPP collaboration is currently operating a 60 L bubble chamber filled with CF$_{3}$I in SNOLAB, Canada.  This target allows for the detection of WIMP-F spin-dependent interactions and WIMP-I spin-independent interactions.  The bubbles formed by particle interactions are observed by two cameras and piezoacoustic sensors.  Rejection of electron recoils is better than 10$^{-10}$.  Alpha particles emitted from the walls of the chamber are a background of concern.  They are identified using acoustic discrimination.  The most recent results from the COUPP collaboration are based on 553 kg-days total exposure of a smaller 4.0 kg CF$_{3}$I detector~\cite{PhysRevD.86.052001}.

\subsection{PICASSO}
The PICASSO experiment is currently operating 32 modular detectors filled with superheated C$_{4}$F$_{10}$ as a target in SNOLAB, Canada.  C$_{4}$F$_{10}$ droplets are suspended in a polymerized gel in a 4.5 L acrylic vessel.  When an incoming particle interacts in the gel, a bubble is formed.  The experiment observes the acoustic deposition by incoming particles with 9 piezoelectric sensors.  The most recent results from the PICASSO experiment are based on a 114 kg-d exposure of 10 modules, containing a total of 0.72 kg of $^{19}$F~\cite{Archambault:2012pm}.

\subsection{SIMPLE}
The Superheated Instrument for Massive ParticLe Experiments (SIMPLE) experiment consists of 15 detectors filled with superheated liquid C$_{2}$CIF$_{5}$ droplets.   The experiment is located in the Laboratorie Souterrain \`{a} Bas Bruit, France.  Bubbles formed by particle interactions are readout by acoustic instrumentation.  The most recent results are based on a science exposure of 18.24 kg-days~\cite{SIMPLE:2014}.

\subsection{Future:  PICO}
The PICASSO and COUPP collaborations have formed a new collaboration called PICO to explore large scale superheated detector options.  The PICO collaboration is currently constructing a 2 L prototype  chamber filled with C$_{3}$F$_{8}$ using the COUPP bubble chamber technology.  Designs are well underway for a larger 250 L experiment is slated for deployment in 2014-2015.

\section{Outlook}

\begin{figure}[th]
\begin{center}
\includegraphics[width=30pc]{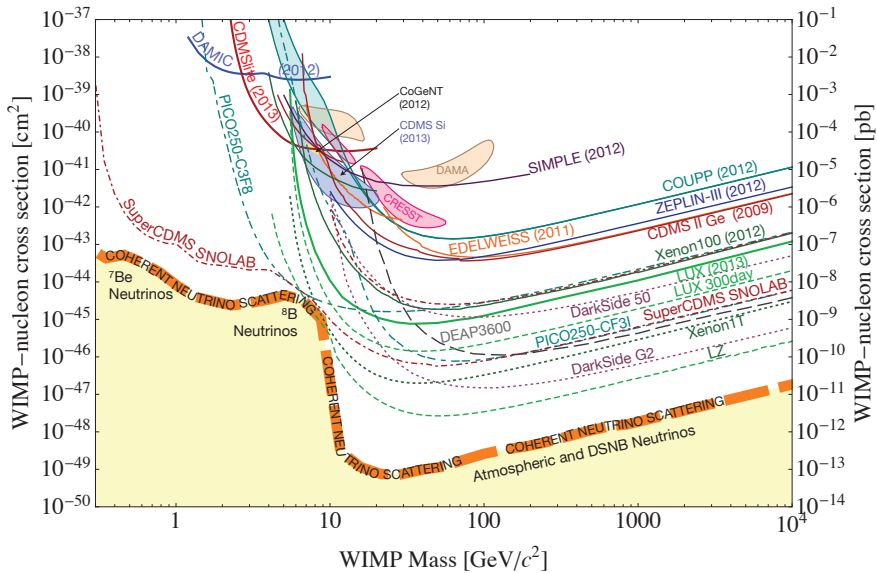}\hspace{2pc}%
\end{center}
\caption{\label{limits} 
A compilation of WIMP-nucleon spin-independent cross section limits (solid lines) and hints of WIMP signals (closed contours) from current dark matter experiments and projections (dashed) for planned direct detection dark matter experiments.  Also shown is an approximate band where neutrino coherent scattering from solar neutrinos, atmospheric neutrinos and diffuse supernova neutrinos will dominate~\cite{Billard:2013qya}.
}
\end{figure}

In the last decade, the experimental dark matter community has made great progress.  We have developed innovative techniques in an effort to suppress backgrounds and extract a dark matter signal in our detectors.  Currently, four experiments have observed excess events.  If these events are interpreted as dark matter CDMS Si and CoGeNT are compatible.  However, it is difficult to reconcile the results from CRESST and DAMA with the null results from other experiments.   At this point, we do not have conclusive evidence of a dark matter signal.  Hence, it is necessary to have experiments using several technologies and a variety of targets located in different locations to maximize the chances of discovery and to confirm any claimed dark matter signal.  Figure \ref{limits} presents the current limits and favored regions of current experiments and projections of the parameter space we will be able to explore with the next generation of experiments.  As we look forward to the next decade, it is clear that with a diverse portfolio we will be able to explore parameter space all the way to the neutrino floor~\cite{Billard:2013qya}.

\section{Acknowledgements}
I would like to acknowledge Frank Calaprice, Juan Collar, Cosmin Deaconu, Klaus Eitel, Graham Giovanetti, Christopher Jackson, Yeongduk Kim, Hugh Lippincott, Eric Miller, Russell Nelson, and Franz Probst who provided me information on the status and the progress of their experimental programs.  I would also like to thank Rob Calkins and Steve Sekula for helpful comments on this article.  This work is supported by the US National Science Foundation Grant Nos. 1151869 and 124640.

\label{}



 \bibliographystyle{elsarticle-num} 
 \bibliography{140320_ref_rev}





\end{document}